\documentclass[aps,prb,twocolumn,superscriptaddress,showpacs,fleqn]{revtex4-1}
\usepackage{color}
\usepackage{graphicx}
\usepackage{amssymb}
\usepackage{dcolumn}
\definecolor{olive}{RGB}{0,128,0}
\usepackage{multirow}

\begin{document}


\title{\textbf{
Plasmonic excitations in quantum-sized sodium nanoparticles studied by time-dependent density functional calculations}}


\author{Jian-Hao Li}
\affiliation{Department of Physics and Center for Theoretical Sciences, National Taiwan University, Taipei 10617, Taiwan}
\affiliation{Center for Condensed Matter Sciences, National Taiwan University, Taipei 10617, Taiwan}
\author{Michitoshi Hayashi}
\affiliation{Center for Condensed Matter Sciences, National Taiwan University, Taipei 10617, Taiwan}
\author{Guang-Yu Guo}\email{gyguo@phys.ntu.edu.tw}
\affiliation{Department of Physics and Center for Theoretical Sciences, National Taiwan University, Taipei 10617, Taiwan}
\affiliation{Graduate Institute of Applied Physics, National Chengchi University, Taipei 11605, Taiwan}


\date{\today}

\begin{abstract}
The plasmonic properties of sphere-like bcc Na nanoclusters ranging from Na$_{15}$ to Na$_{331}$ have been
studied by real-time time-dependent local density approximation calculations.
The optical absorption spectrum,  density response function and static polarizability are evaluated.
It is shown that the effect of the ionic background (ionic species and lattice)
of the clusters accounts for the remaining discrepancy
in the principal (surface plasmon) absorption peak energy between the experiments and previous calculations based
on a jellium background model. The ionic background effect also pushes the critical cluster size where the maximum width of
the principal peak occurs from Na$_{40}$ predicted by the previous 
jellium model calculations to Na$_{65}$. In the volume mode clusters
(Na$_{27}$, Na$_{51}$, Na$_{65}$, Na$_{89}$ and Na$_{113}$) in which the density response function
is dominated by an intense volume mode, a multiple absorption peak structure also appears next to the principal peak.
In contrast, the surface mode clusters of greater size (Na$_{169}$, Na$_{229}$, Na$_{283}$ and Na$_{331}$)
exhibit a smoother and narrower principal absorption peak because
their surface plasmon energy is located well within that of the unperturbed electron-hole
transitions, and their density responses already bear resemblance to that of classical Mie theory.
Moreover, it is found that the volume plasmon that exist only in finite size particles, gives rise to
the long absorption tail in the UV region.
This volume plasmon manifests itself in the absorption spectrum even for clusters as large as Na$_{331}$
with an effective diameter of $\sim$3.0 nm.
\end{abstract}

\maketitle
\section{Introduction}
The prominent outlook of plasmonics for various applications in nanotechnology has stimulated its intense
studies~\cite{Mai07, HLC11, ZPN09, MdN11, SKD12}, notwithstanding
the plasmon, a collective type of electronic excitations generally featured by metallic systems, has been an
old topic in physics.  Past theoretical and experimental investigations on plasmons in metal nonoclusters
have mainly been focused on the Mie surface plasmons~\cite{Mie8}, as opposed to volume plasmons which cannot be
optically excited in bulk metals~\cite{Fer58}.  However, volume plasmons in small metal
nanoclusters have been theoretically shown to be non-negligible and were also experimentally demonstrated
recently~\cite{Kre90, XYK09, Bra93, BWM08}.  Inspired by these pioneering works, therefore, it is
interesting to further examine the plasmonic properties of finite metallic particles.  Furthermore, in
subnanoscale size regime quantum plasmons can have distinct behaviors that are different from classical
plasmons~\cite{YG08, ZPN09, MdN11, HLC11, SKD12} due to, for example, Landau
fragmentation~\cite{Yan93, Yan98, NKR02}, and hence call for further investigations.  In this paper, we
focus on quantum-sized neutral Na metal clusters with the number ($N$) of Na atoms ranging from $N=15$ to
$N=331$, and use quantum mechanical time-dependent (TD) local density approximation
(LDA)~\cite{RG84, MUN06, Sla74, VWN80} to study how the optical absorption spectrum, its related
properties, and linear density response function of the clusters evolve as the cluster size increases.  In
particular, by examining the spatial distribution of density response function for different frequencies, we
investigate how the compositions of both surface and volume plasmons depend on the cluster size.  The
alkali metal, Na, is chosen because it bears a wide range of absorption energy dominated by the plasmonic
excitations without interference from the deeper electrons such as $d$-states in transition metals to
contribute to interband transitions~\cite{JC72, PNH03, YG08}.

In the present TDLDA calculations, we employ the real time and real space propagation
method~\cite{YB96, YG08} in the linear response regime where the two properties of interest, absorption
spectrum and density response function to uniform perturbation, can readily be extracted.  This approach
is on the same theoretical level as the one employing Dyson-type equation for the density-density response
function within TDLDA~\cite{ZS80, Eka85}, or the matrix TDLDA method~\cite{C95}, which constructs 
a random phase approximation (RPA)-like matrix equation.
In order to study the Landau fragmentation
of the surface plasmon, we also calculate for several clusters the oscillator strength of unperturbed
electron-hole transitions that are used as the basis set in the matrix TDLDA method.  For studying plasmonic
properties of metallic particles, there have also been such works~\cite{Bra93} as
employing RPA-like matrix equations~\cite{Yan93, Yan98, NKR02}, to formulating Dyson-type equation for
density-density response function~\cite{Eka85} or effective electron-electron-interaction~\cite{Kre90}.

For metal nanoclusters, one of the outstanding characteristics has been the electronic shell structure
~\cite{Kni84,Bra93,dH93} in which a cluster is stabilized by forming shell closure at certain numbers ($N_{e}$) of
(nearly) free valence electrons, i.e., $N_{e} = 8, 20, 40$ and so on.
This reflects an effective spherical potential these electrons experience, and, therefore, several theoretical works have been based on the
spherical jellium model or its variants with considerable success to describe the positive background of a
metal cluster~\cite{Bra89, LRM91, Yan93, Bra93, MR95, Yan98, NKR02, Bec91, MG95}.  For a more precise
prediction on the nanocluster properties, however, taking the real ionic background
(i.e., ionic species and lattice) into account explicitly
should be important.  This is true especially for small nanoclusters in low temperatures or fixed in
space as in nano-device, where the effect of discreteness of the ionic background on the observables cannot
be averaged out.  For example, it has been shown~\cite{CRS98} that the presence of ionic background can
cause more spectral splittings than a corresponding jellium model.  Moreover, the discrepancy~\cite{WIJ06}
between experiment and jellium model calculation in the principal (surface plasmon) absorption peak energy
of free metallic particles remains to be resolved.
Unlike the spectral splitting, the principal peak energy is relatively insensitive to
temperature~\cite{HB04, SEK99}.

If one aims to take into account the real ionic structure of a metal cluster, however, the construction of
the ground state structure often relies on numerous configuration searches~\cite{KBR00, SSG02} whose
complexity rapidly arises as the cluster size increases.  Furthermore, there is no guarantee that the ground state
structure or other important isomers would be located; different theoretical calculations can also predict
different energy orders for the same set of structures.  Hence, instead of searching from many geometries
with magic electron numbers, in order to build Na clusters, we simply pile up atoms layer by layer onto
bulk Na bcc lattice and this helps construct a series of highly symmetric Na clusters ranging 
from $N = 1$ to as large as $N = 331$ (Table I).
The structures of these Na clusters are then theoretically optimized and used in the present TDLDA calculations. 
Since a plasmon is formed by the collective motion of (nearly)
free electrons (see, e.g., Ref.~\onlinecite{RS04} for a review), it is assumed that the general trends
of plasmonic properties of metal clusters are more susceptible to the overall shape, size and electronic
density than to the exact atomic structures.  In addition, the conventional
Hill-Wheeler quadrupole deformation parameters, $\beta_{2}$, and the octupole moment,
$\beta_{3}$,~\cite{YL94, MR95} for the constructed clusters are all equal to zero.  Therefore, possibility
of spectral splitting due to particle shape deformation could be excluded.  Moreover, Yannouleas and
Landman~\cite{YL94} showed that as the cluster size increases, the deformation of the stable structures
from a spherical shape generally diminishes. Most clusters studied here are larger than $N = 60$,
whose deformations were found to be very small~\cite{YL94}.  The generated highly symmetric clusters also
facilitate numerical calculations and help provide systematic knowledge on the effect of size evolution on
the plasmonic properties of metal nanoclusters.

The paper is organized as follows.  In Sec. II, computational procedures are introduced, including the
construction of various Na metal clusters and the details of the present TDLDA calculations.  In Sec.
III, the calculated optical absorption spectra and the related properties are examined.  In Sec. IV, the
density response function of each cluster for each major absorption feature is analyzed.
Finally, conclusions are given in Sec. V.

\section{Theory and computational details}
Starting from a single Na atom, a series of nearly spherical nanoclusters are constructed by attaching one
new layer of atoms selected from the Na bcc lattice~\cite{AM76} a time to the previous smaller cluster.
Each layer has atoms of equal distance to the cluster center.
An effective radius (\emph{R$_{C}$}) for measuring the size of a cluster is defined by
$(4\pi/3)R_{C}^{3}=NV_{c}$ where $V_{c}$ is the volume per atom of the bulk Na. The details of all the
constructed Na clusters are listed in Table I.

\begin{table}
\caption{
Details of the Na nanoclusters used as the initial conformations for structural optimization in this paper.
Na atoms on the bcc lattice are divided into shells (layers) where shell 0, a single Na atom, is at the
center upon which other shells are stacked one by one to make up various clusters of different sizes
considered here.  In each shell, only the atom(s) at the first octant are listed.  Other atoms on the shell
can be obtained by applying point group operations to these atoms.  \emph{R$_{L}$} stands for the shell
distance to the cluster center, whereas \emph{R$_{C}$} is the effective cluster radius (see text).}
\begin{ruledtabular}
\begin{tabular}{ccccccc}
 Shell&Cluster &   $x$   &   $y$   &   $z$   & \emph{R$_{L}$}(\AA) & \emph{R$_{C}$}(\AA) \\ \hline
 0  &          &     0 &   0   &  0  &   0.00 &     \\
 1  &          &     $a/2$ &   $a/2$   &  $a/2$  &   3.66 &     \\
 2  &  Na$_{15}$ &    $a$   &   0   &  0  &   4.23 &  5.13  \\
 3  &  Na$_{27}$ &    $a$   &   $a$     &  0  &   5.98 &  6.24  \\
 4  &  Na$_{51}$ &    $3a/2$ &   $a/2$   &  $a/2$  &   7.01 &  7.72  \\
 5  &          &     $a$   &   $a$     &  $a$    &   7.33 &        \\
 6  &  Na$_{65}$ &    $2a$   &   0   &  0  &   8.46 &   8.37  \\
 7  &  Na$_{89}$ &    $3a/2$ &  $3a/2$   &  $a/2$  &   9.22 &   9.29  \\
 8  &  Na$_{113}$&    $2a$   &   $a$     &  0  &   9.46 &  10.06  \\
 9  &          &     $2a$   &   $a$     &  $a$    &  10.36 &        \\
 10  & Na$_{169}$&    $3a/2$  &  $3a/2$   & $3a/2$  &  10.99 & 11.50 \\
     &         &   $5a/2$  &   $a/2$   &  $a/2$  &    &    \\
 11  &          &    $2a$    &  $2a$     &  0  &  11.96 &       \\
 12  &  Na$_{229}$&   $5a/2$  &  $3a/2$   &  $a/2$  &  12.51 & 12.73  \\
 13  &          &    $2a$    &  $2a$     &  $a$    &  12.69 &       \\
     &          &    $3a$    &  0    &  0  &       &      \\
 14  &  Na$_{283}$&   $3a$    &   $a$     &  0  &  13.38 & 13.66  \\
 15  &          &   $5a/2$  & $ 3a/2$   & $3a/2$  &  13.87 &        \\
 16  &  Na$_{331}$&   $3a$    &  $a$     &  $a$    &  14.03 & 14.39  \\
\end{tabular}
\end{ruledtabular}
\end{table}

The structures of these constructed clusters are then theoretically optimized.
The structural optimizations are performed within the density functional theory
with the generalized gradient approximation~\cite{PBE96} by using the accurate 
projector-augmented wave method, as implemented in the VASP package~\cite{KH9394, KF96, KFu96}.
The shallow semicore 2$p$ electrons are treated as the band states.
The cut-off energy for plane waves is 300 eV. The lattice constant of bulk Na determined this way 
agrees with the experimental value within 0.2 \%.  We find that after optimization, all
the clusters listed in Table I remain in the high cubic $O_h$ symmetry. Nevertheless,
the distances of the atoms on the surface of each cluster can vary up to a few percents
compared to that of structurally unrelaxed clusters, pushing the overall shape of the clusters
towards a spherical one.

The optical absorption spectra of these optimized clusters are then calculated by real time and real space TDLDA, as
implemented in the Octopus package~\cite{MCB03}.  Each cluster is positioned in a large box built by
surrounding a 10 \AA\ sphere around each atom and putting into use the overlapping region.
The spacing of the grid used for dividing the spatial region is 0.3 \AA. Further calculations
for the benchmark system Na$_{9}$ using a surrounding sphere of a 15 \AA$ $ radius
and a grid spacing of 0.15 \AA$ $, indicate that the radius of
surrounding spheres and the grid spacing employed here have rendered the calculated ground state energy
and the principal peak energy in the absorption spectrum converged within 0.02 \%
and $\sim$0.01 eV, respectively.

Only the valence 3$s$ electrons responsible for the plasmonic oscillation are treated by the TDLDA.  The
ionic core of the Na atoms is taken into account by using a Troullier-Martins
norm-conserving pseudopotential~\cite{TM91} in non-local form~\cite{KB82}. Moreover, since all the
studied clusters have odd number of atoms, spin unrestricted calculations are performed.  Nonetheless,
the electronic structure of these clusters has been
restricted to the lowest spin configuration, $\Delta_{\alpha\beta}=1$, in the present LDA calculations, where
$\Delta_{\alpha\beta}$ denotes the number difference between spin up and spin down electrons.  This
should be reasonable since natural clusters generally have the lowest net spin configuration,
especially for large clusters~\cite{ARS08}.

To calculate the optical absorption spectrum of each cluster using real time/space TDLDA, we follow the
spectral analysis procedure proposed by Yabana and Bertsch~\cite{YB96} by using a delta-impulse of
electric field,
\begin{equation}
\mathbf{E}=K\delta(t)\hat{x},
\end{equation}
where $K$ should be small enough to stay in the linear response regime, to excite all frequencies of the system in its
ground state.  This is achieved by setting
\begin{equation}
\varphi_{i}(\mathbf{r},\delta t)=e^{iKx}\varphi_{i}(\mathbf{r},0)
\end{equation}
where $\varphi_{i}(\mathbf{r},0)$ are the ground state Kohn-Sham orbitals.  The system is then
propagated for a certain time, which equates propagating the Kohn-Sham orbitals directly,
\begin{equation}
\varphi_{i}(\mathbf{r},t+\Delta t)=U(t+\Delta t,t)\varphi_{i}(\mathbf{r},t),
\end{equation}
where the time-reversal-symmetry propagator
\begin{equation}
U(t+\Delta t,t)=\textrm{exp}\{-i\frac{\Delta t}{2}H(t+\Delta t)\}\textrm{exp}\{-i\frac{\Delta t}{2}H(t)\}.
\end{equation}
\noindent
The total propagation time used is $T = 50$ $\hbar$/eV, with each time step being 0.005 $\hbar$/eV.
After propagation, the dipole oscillation strength function per atom,
\begin{equation}
S_{xx}(\omega)=\frac{2m_{e}\omega}{\pi e^{2}NK}\textrm{Im}\int_{0}^{T}dt e^{i\omega t-\Gamma t}[\mu_{x}(t)-\mu_{x}(0)]
\end{equation}
\noindent
is calculated, where $\mu_{x}(t)$ is the dipole moment along the $x$-axis of the system.
$\Gamma$ is the damping factor which is
set to $0.14$ eV/$\hbar$ to simulate the intrinsic damping due to the finite temperature in experiment other
than Landau fragmentation~\cite{Yan93}.  Using a normalized signal function of  
$\frac{2}{\pi} \textrm{sin}(\omega_{0}t)$ to replace the 
[$\mu_{x}(t)-\mu_{x}(0)$] in Eq. (5), it can be shown that the finite propagation 
time $T = 50$ $\hbar$/eV used here is able to reproduce the exact $S_{xx}$ ($T = \infty$ $\hbar$/eV) 
within $\pm 0.005$ eV$^{-1}$ at the damping factor $\Gamma = 0.14$ eV/$\hbar$.
Because of the symmetry of the clusters, only the calculations with
electric field applied along one Cartesian axis are necessary.  Thus,
\begin{equation}
S(\omega)=\frac{1}{3}\sum_{i=1}^{3}S_{ii}(\omega) \simeq S_{xx}(\omega)
\end{equation}
is eventually the calculated absorption spectrum
that satisfies the Thomas-Reiche-Kuhn (TRK) sum rule~\cite{BLM79},
\begin{equation}
\int_{0}^{\infty}d\omega S(\omega)=1.
\end{equation}

In order to study the formation process of the absorption profiles near the principal absorption peak that
are mediated by Landau fragmentation~\cite{Yan93}, we also examine for several clusters the couplings among
the unperturbed individual electron-hole transitions that are used as the basis set for diagonalization in
matrix TDLDA method~\cite{C95}.  The oscillator strength of an unperturbed electron-hole transition is given
by
\begin{equation}
f_{I}=\frac{2m}{3\hbar}\omega_{I}\sum_{i=1}^{3}|\langle\Psi_{0}|\hat{x_{i}}|\Psi_{I}\rangle|^{2}
\end{equation}
\noindent
where $\langle\Psi_{0}|\hat{x_{i}}|\Psi_{I}\rangle$ is the transition dipole moment between the ground and
electron-hole excited state along $x_{i}$ direction and $\omega_{I}$ the energy difference between the
electron and hole Kohn-Sham orbitals. Note that a principal absorption
peak is formed by all the couplings among individual electron-hole transitions.  The corresponding charge
density oscillation may therefore contain the surface section, which is naturally the surface plasmon
component, and volume section, which is from the coupled individual transitions.  This
will be discussed in more detail in Sec. IV.  Thus, we call the combined excitation the
"Landau fragmented surface plasmon", whereas "surface plasmon" refers only to its surface section.  The
former can be seen as formed by the coupling between the latter and the nearly degenerate
electron-hole transitions.  It is of course difficult to clearly separate the two contributions, but the
surface plasmon can be assigned as formed by the couplings among all the electron-hole
transitions excluding those that fall within $\omega_{pp}\pm\eta$, where $\omega_{pp}$ denotes the principal
absorption peak energy.  The surface plasmon energy is generally very close to $\omega_{pp}$, as shown in
Ref.~\onlinecite{Yan93}, where it is called before-breakup surface plasmon.

On the other hand, for calculating the linear density response function ($\widetilde{\chi}(\mathbf{r},\omega)$)
of a system in the ground state to a constant (in spatial and frequency domain) external potential,
we perform a similar finite time Fourier transform to the induced electronic density in a TDLDA
calculation, i.e.,
\begin{equation}
\widetilde{\chi}(\mathbf{r},\omega)=\frac{1}{K}\textrm{Im}\int_{0}^{T}dt e^{i\omega t-\Gamma t}[n(\mathbf{r},t)-n_{0}(\mathbf{r})]
\end{equation}
\noindent
where $n_{0}(\mathbf{r})$ and $n(\mathbf{r},t)$ are the electronic density at the ground state (initial
time) and at time $t$, respectively.

After obtaining the $\widetilde{\chi}(\mathbf{r},\omega)$, the volume density response proportion (VRP) at
a frequency is estimated through the computation of
$\int_{in} |\widetilde{\chi}(\mathbf{r},\omega) |d^{3}r$ and $\int_{out} |\widetilde{\chi}(\mathbf{r},\omega)|d^{3}r$.  
The division of the inner and outer regions of a cluster is
achieved by employing the $0.0250$ e/\AA$^{3}$ isosurface on the ground state electronic density at the
border between the cluster and vacuum.  While the average valence electron density inside the bulk Na is
$0.0265$ e/\AA$^{3}$ (Ref. ~\onlinecite{AM76}), the density fluctuation inside a cluster can result in the
inter-atomic density being lower than this value.  The $0.0250$ e/\AA$^{3}$, on the other hand, is found
to be well below this fluctuation and is accordingly used for defining the cluster borders
beyond which electronic density continues decaying towards the vacuum.  The defined boundary for each
cluster on the $xy$-plane which slices through the cluster center, is shown in Fig. 1.

\begin{figure}
\raggedright
\begin{center}
\includegraphics[angle=0,width=8.0cm]{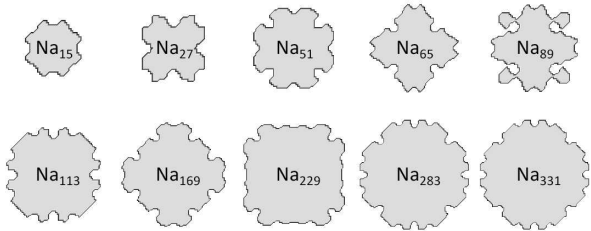}
\end{center}
\caption{Boundary of the clusters studied here: the boundary contour on the $xy$-plane that slices 
through the center of each cluster. The boundary is used to define the volume region of a cluster (see text).}
\end{figure}

To measure the VRP value at a given frequency $\omega$,
\begin{equation}
\int_{in}|\widetilde{\chi}(\mathbf{r},\omega)|d^{3}r/\int_{in+out}|\widetilde{\chi}(\mathbf{r},\omega)|d^{3}r
\end{equation}
is then calculated.  By further referencing the distribution of $\widetilde{\chi}(\mathbf{r},\omega)$ on
the $xy$-plane for each feature in an optical absorption spectrum or VRP spectrum, the properties of the
plasmonic or other excitations can be studied in detail.

\section{Optical absorption spectra}
\subsection{Oscillation strength and peak energy}
\begin{figure}
\raggedright
\begin{center}
\includegraphics[angle=0,width=8cm]{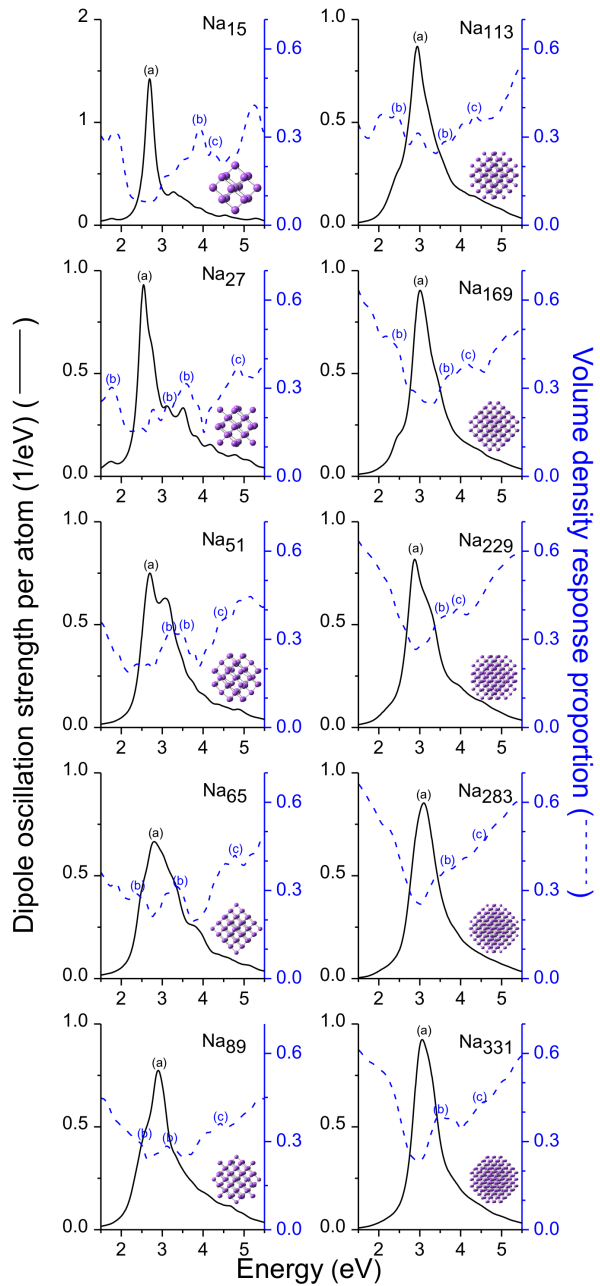}
\end{center}
\caption{(Color online) Calculated absorption spectra (dipole oscillation strength) of the Na
clusters studied here.  Also plotted are the volume density response
proportion (VRP) spectra.}
\end{figure}

\begin{figure}
\raggedright
\includegraphics[angle=0,width=8cm]{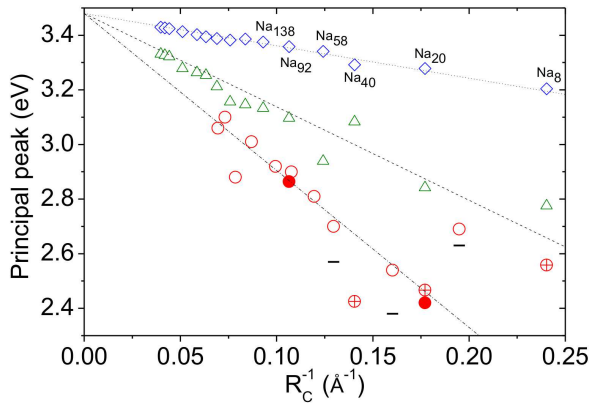}
\caption{(Color online) Calculated principal peak energy ({\color{red}$\circ$})
vs. the reciprocal of the effective Na cluster radius.  The TDLDA results for three structurally 
unoptimized clusters (see Table I) are also shown ($-$).
The results from previous LDA/electron spill-out model ({\color{blue}$\diamond$})~\cite{WIJ06}
and TDLDA ({\color{olive}$\triangle$})~\cite{WIJ06} calculations based on a spherical jellium
model, as well as the experimental data ({\color{red}$\oplus$}~\cite{SK91} 
and {\color{red}$\bullet$}~\cite{XYK09}), are also displayed for comparison.
Dotted, dashed, and dash-dotted lines, which converge to $\omega_{Mie}$, are a guide to the eye only, for the open diamonds, open
triangles, and all types of circles, respectively.  While the coupling of the surface plasmon with
electron-hole transitions leads to the red-shift (dotted to dashed line) of
the principal peak energies~\cite{WIJ06}, the ionic effect is evidently responsible for the further red-shift
(dashed to dash-dotted line) that brings our results in good agreement with the experiments.}
\end{figure}

The calculated optical absorption and VRP spectra of each cluster considered here in the energy
range 1.5-5.5 eV are plotted in Fig. 2.  An important feature to be examined first is the principal
(strongest) peak position which corresponds to the surface plasmon.  Fig. 3 shows the principal peak
energies of our Na clusters and also of four closed-shell Na clusters (Na$_{8}$, Na$_{20}$, Na$_{40}$,
and Na$_{92}$) measured before~\cite{SK91,XYK09}, together with the results of previous LDA/electron
spill-out model and TDLDA/jellium background calculations for several closed-shell clusters~\cite{WIJ06}.
Note that the measured principal peak energies for closed-shell clusters (hence without deformation effect)
are much smaller than the Mie frequency~\cite{Bra93,dH93,RS04},
${\omega}_{Mie}=\omega_{p}/\sqrt{3}$,
where $\omega_{p}$ is the bulk plasmon frequency.  The discrepancy has usually been attributed to the
electron spill-out on the surface~\cite{Bra93, RESH95, RS04}.  If the electron spill-out effect is taken
into account, the surface plasmon energy becomes
${\omega '}_{Mie}^{2}=\omega_{Mie}^{2}(1-\Delta N_{e}/N_{e})$,
where $\Delta N_{e}$ is the spilled-out electrons.  However, as can be seen from the discrepancies between the
results of the LDA/spill-out model and experimental data in Fig. 3, the principal peak energies remain notably off the
experimental trend.  Note that the theoretical trend can be further corrected if one considers the coupling
of the surface plasmon with the individual electron-hole excitations~\cite{WIJ06,GG02}.
As shown in Fig. 3, the TDLDA results move closer toward the measured data.  Despite the
improved description, however, the remaining discrepancy (see Fig. 3) is not yet clearly understood,
not least the early energy decrease with increasing cluster size of Na$_{8}$, Na$_{20}$, and Na$_{40}$ that is
not predicted by either LDA/spill-out model or TDLDA/jellium background calculations.  Note that the energy
shift due to the finite temperature in experiments is much smaller and can be ignored in this
context~\cite{HB04, SEK99}.  A natural conjecture would be that the errors in the jellium approximation of the positive
background could account for the remaining deviation.  Indeed, K\"{u}mmel \emph{et al.}~\cite{KBR00} showed
that for clusters of up to 56 electrons, the explicit inclusion of ionic structure can considerably improve
the agreement with experiments.  However, a theoretical confirmation of the ionic effect on a larger size
scale is still needed, which becomes extremely difficult for the search of ground state and isomer
structures.  Nevertheless, with the relatively simple cluster construction scheme employed in this work,
the ionic effect for much larger clusters can be investigated.

It can be seen from Fig. 3 that the experimental trend is well reproduced by our TDLDA results, i.e.,
the principal peak energies follow a line that meets the experimental data at Na$_{20}$ and Na$_{92}$
(which appears to be the largest neutral cluster ever measured).  The decrease of peak energies from
Na$_{15}$ to Na$_{27}$ also agrees with the experimental results for Na$_{8}$, Na$_{20}$, and Na$_{40}$.
The energy plummeting at Na$_{229}$ may be due to an incidental strong coupling of some nearly degenerate
electron-hole transitions to the surface plasmon.  We have also performed TDLDA calculations for the
structurally non-optimized clusters (Table I) and the principal peak energies are further red-shifted (Fig. 3).  
Since all the clusters considered here are with a similar average background density and a similar global shape to the
corresponding jellium spheres, the main factor responsible for the overall red-shift of the principal peak
energy and the red-shift with size for small clusters should be the real ionic background of these clusters.
Note that for the non-optimized clusters that have the same background density as the jellium spheres, the
principal peak energies further red-shift away from (rather than move closer to) the jellium results, clearly indicating that
the presence of the realistic ionic background would lower the principal peak energy.  This also suggests
that the overall denser and non-uniform background density resulted from the structural relaxation, which
is not taken into account by the jellium model, is also an important factor for the energy shift of the
surface plasmon.  While the red-shift due to the use of the realistic ionic background could be accounted 
for to some extent by the softened boundary potential which could be modeled using a "soft" jellium background~\cite{RS04},
the structural relaxation may need to be taken into account properly in order to better reproduce 
the experimental trend.

\begin{figure}
\raggedright
\includegraphics[angle=0,width=8cm]{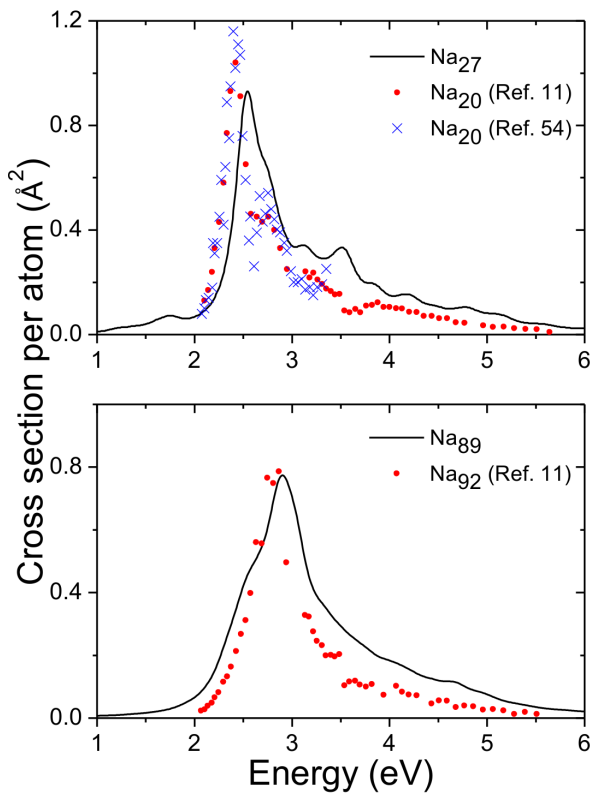}
\caption{(Color online) Calculated optical absorption cross-sections of Na$_{27}$ (Na$_{89}$)
compared to the measured spectra~\cite{XYK09} of Na$_{20}$ (Na$_{92}$).  
Measured data of Na$_{20}$ (crosses) from a different work~\cite{PWK91} are also plotted.
The good agreement between the calculations and experiments indicates that the plasmonic properties
depend more strongly on the overall shape, size, and ionic density of a cluster than on its detailed atomic
structure.}
\end{figure}

For a detailed comparison, the measured absorption cross section per atom [$\sigma(\omega)$] of
Na$_{20}$ and Na$_{92}$~\cite{XYK09} together with our calculated ones of Na$_{27}$ and Na$_{89}$
are displayed in Fig. 4.  $\sigma(\omega)$ relates to the dipole oscillation strength function,
$S(\omega)$, by
$\sigma(\omega) = (\pi e^{2}\hbar/2\epsilon_{0}mc)S(\omega)$,
where $c$ is the speed of light, and thus satisfies the TRK sum rule
\begin{equation}
\int_{0}^{\infty}\sigma({\omega})d\omega=1.0975 eV\cdot\AA^2.
\end{equation}
It can be seen that the main features in the measured spectrum of Na$_{20}$ (Na$_{92}$) and theoretical
spectrum of Na$_{27}$ (Na$_{89}$) are very similar; apart from the close principal peak positions, 
there are bumpy structures between 2.5 (3.0) eV and 5.0 eV for
Na$_{20}$/Na$_{27}$ (Na$_{92}$/Na$_{89}$) and a long tail extending up to at least 5.5 eV.


\subsection{Static polarizability}

\begin{figure}
\raggedright
\includegraphics[angle=0,width=8cm]{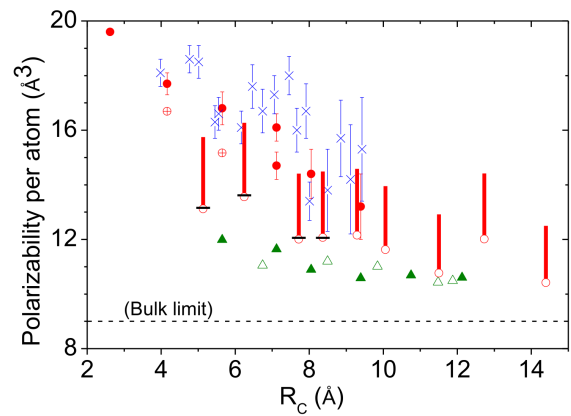}
\caption{(Color online) Calculated static polarizability ({\color{red}$\circ$}) vs. the effective Na cluster radius,
with a +20 \% bar accounting for possible temperature effect up to 500 K being added.  The results ($\mathbf{-}$) 
obtained using Sternheimer equation within TDDFT formalism\cite{ABMR07} for some clusters are also plotted.
For comparison, experimental data for closed-shell clusters ({\color{red}$\bullet$})~\cite{TK01}
(with error bar present) and ({\color{red}$\oplus$})~\cite{RA99} are also plotted, together with
previous measurements on non-closed-shell clusters ({\color{blue}$\times$})~\cite{TK01} and theoretical
results for closed-shell ({\color{olive}$\blacktriangle$}) and non-closed-shell ({\color{olive}$\triangle$})
clusters from the TDLDA/jellium background calculations~\cite{Eka85}.}
\end{figure}

Note also that the measured absorption spectrum of Na$_{20}$ and Na$_{92}$ has not exhausted the TRK
sum rule (71\% for Na$_{20}$ and 60\% for Na$_{92}$) and this was attributed to the yet unfound
photoabsorption channels in other frequency ranges~\cite{XYK09}.  However, by computing the
underlying area of each absorption spectrum between 2.0 and 5.5 eV, we find that for all the
clusters considered, 87\% or higher percentage of the oscillator strength has been accounted for.
We further examine our results by evaluating
the static polarizability ($\alpha$) by using the sum rule
\begin{equation}
\alpha=\frac{e^{2}}{4\pi\epsilon_{0}m_{e}}\int_{0}^{\infty}\frac{S(\omega)}{\omega^{2}}d\omega.
\end{equation}
The $\alpha$ of several clusters have also been calculated using Sternheimer equation within TDDFT
formalism~\cite{ABMR07} and agree excellently with the results obtained from the sum rule.
Our calculated $\alpha$ are compared with the experiments on several closed-shell clusters in Fig. 5.
Measured data of several non-closed-shell clusters are also plotted as a reference to the electronic shell structure effect on
$\alpha$ that will be discussed below.  K\"{u}mmel \emph{et al.}~\cite{KAM00} demonstrated that thermal
expansion coefficient for small metal clusters is considerably larger than that in the bulk, which in turn
leads to a substantial increase of $\alpha$\ at room temperature that can account for the long-standing
discrepancy between theories and experiments.  Blundell \emph{et al.}~\cite{BGZ00}, studying temperature
effect as well, also showed that closed-shell Na clusters ranging from 8 to 139 electrons at 300 K all
exhibit a roughly 15 \% larger $\alpha$ than at 0 K.  From the linear relation between temperature and
average $\alpha$ for Na$_{8}$ and Na$_{10}$~\cite{KAM00}, it can be deduced that at 500 K (a typical
temperature in experiments) the enhancement of $\alpha$ is around 20\%.  We therefore add a +20 \% bar to
our results to simulate the possible effect of temperature up to 500 K.  Note that the precise relation
between the increase of $\alpha$ and temperature (and size~\cite{KAM00}) still waits for further
clarification.  Note also that due to various factors such as temperature difference or isomerism
effects~\cite{TK01}, the experimental data exhibit uncertainties as large as 20-30 \%, as shown in
Fig. 5.  Nevertheless, in the size regime where measured data are available, our
calculated static polarizabilities are in reasonable overall agreement with the experiments.  The oscillation
of $\alpha$ as a function of cluster size not observed in experiments on closed-shell clusters, may be due to
non-closed electronic shell that can lead to the oscillation of $\alpha$, either indirectly from the cluster
deformation, or directly as exhibited by our results.  The latter mechanism can also be seen (Fig. 5) in
the results of the TDLDA/jellium background calculations~\cite{Eka85} for several non-closed shell clusters
where an oscillation as a function of cluster size is observed.  The difference between the
TDLDA/jellium background results and ours should be due to the real ionic structure
used in the present calculations.

\begin{figure}
\raggedright
\includegraphics[angle=0,width=8cm]{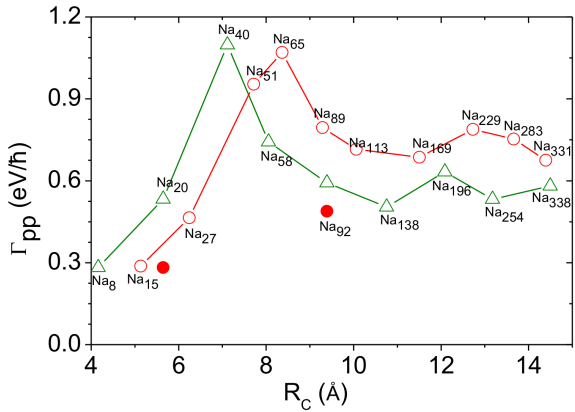}
\caption{(Color online) The widths (FWHMs) ({\color{red}$\circ$}) of the principal peaks
derived from the absorption spectra from the present work (Fig. 2) vs. the effective Na cluster radius.
Also plotted are the FWHMs from previous matrix RPA-LDA/jellium model calculations
({\color{olive}$\triangle$})~\cite{Yan93} and the experimental widths
({\color{red}$\bullet$}) of Na$_{20}$ and Na$_{92}$~\cite{XYK09} derived from the absorption spectra
shown in Fig. 4.}
\end{figure}

The good agreement of our optical absorption spectrum and static polarizability with the experiments
thus indicates that the contribution from other photoabsorption channels outside the energy range 2.0-5.5
eV accounts for $\sim$13 \% or less, and our calculated absorption spectra of Na$_{27}$ and Na$_{89}$ provide
possible candidates for missing absorptions of Na$_{20}$ and Na$_{92}$, respectively (Fig. 4).  Since, as
indicated by the authors of Ref. ~\onlinecite{XYK09}, there is a 15 \% uncertainty in the measured data for Na$_{20}$ and
a 20 \% for Na$_{92}$, it is likely that the difference is caused by the remaining uncertainty
of the experimental data.  For example, Fig. 4 also shows the measurements on Na$_{20}$ in
which the principal peak is higher and hence can increase the underlying cross section area.

\begin{figure}
\raggedright
\includegraphics[angle=0,width=8cm]{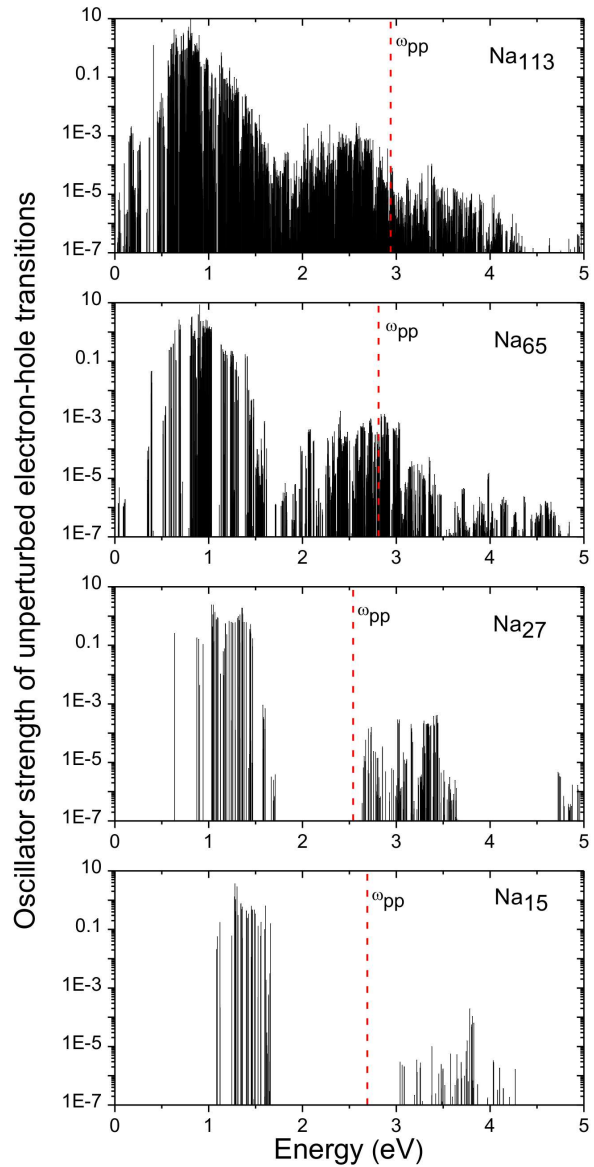}
\caption{(Color online) Oscillator strengths of the unperturbed electron-hole transitions that form the basis for matrix
TDLDA diagonalization.  To examine four main stages of Landau fragmentation, 40, 70, 165, and 285 lowest
energy Kohn-Sham $\alpha$-spin states, plus the same amount of lowest energy $\beta$-spin states, for
Na$_{15}$, Na$_{27}$, Na$_{65}$, and Na$_{113}$, respectively, are used to calculate these transitions.
Na$_{15}$: before fragmentation, as indicated by a narrow and smooth principal peak
(Fig. 2); Na$_{27}$: fragmentation going to start, as indicated by a rough absorption profile on the right
side of the principal peak; Na$_{65}$: fragmentation at its maximum, as revealed by the maximum width of the
principal peak (Fig. 6); Na$_{113}$: fragmentation close to a local minimum, indicated by the shrinking
width around a local minimum.  The principal peak energy ($\omega_{pp}$) for each cluster is marked by the
red dashed line.}
\end{figure}

\subsection{Spectral width}
Now let us move onto the width of the principal absorption peak, which is estimated by the
full-width-at-half-maximum (FWHM).  Fig. 6 shows the results of our studied clusters derived from Fig. 2,
compared with those of the previous matrix RPA-LDA/jellium background calculations for several closed-shell
clusters~\cite{Yan93} and also of the experiments for Na$_{20}$ and Na$_{92}$ (Fig. 4).  
Note that the damping factor of $0.14$ eV/$\hbar$ used in the present work has rendered the width of Na$_{15}$ matches 
well with that of Na$_{8}$ from the matrix RPA-LDA/jellium model calculation and the experimental result of Na$_{20}$.  For
Na$_{20}$ or smaller clusters the Landau fragmentation should not start yet, as evidenced from the
simple shape of the principal peak of Na$_{20}$ that is well separated from the bumpy absorption profile on the
higher-energy side by $\sim$0.2 eV (Fig. 4).  Hence, the width should reflect the intrinsic damping factor due
to the finite temperature in experiment~\cite{Yan93}.  More details will be given in the following text.

The complicated absorption profile near the principal peak in most clusters (Figs. 2 and 4) should
be largely due to the individual electron-hole transitions, as mentioned before.  It is
therefore desirable to further study the formation process of these profiles in more detail.  For Na$_{15}$,
Na$_{27}$, Na$_{65}$, and Na$_{113}$, the oscillator strengths [Eq. (8)] of the unperturbed electron-hole
transitions used in the matrix TDLDA method~\cite{C95} are plotted in Fig. 7.  Note that the matrix TDLDA
method differs from the matrix RPA-LDA~\cite{Yan93} in that the latter employs semiempirical single-particle
potentials.

It was found~\cite{Yan93} that Na$_{20}$ is the cluster of critical size for which the surface plasmon
starts to be fragmented due to the energy gap closing between the surface plasmon and the unperturbed
electron-hole transitions.  For clusters smaller than Na$_{20}$, the surface plasmon energy lies within
the gap between $\Delta n=1$ and $\Delta n\geq3$ electron-hole transitions, where $n$ denotes the principal
quantum number of single-particle electronic levels in a jellium sphere~\cite{Yan93}.
As the cluster size increases, the gap narrows due to the increase of the cluster volume, until the edge at
$\Delta n\geq3$ side finally touches the surface plasmon energy, leading to a strong coupling and thus
Landau fragmentation of the surface plasmon.  For larger clusters, the surface plasmon energy continues to
stay in the $\Delta n\geq3$ transition forest, in which the small gaps between the transitions are due to
the flattened potential outside the cluster.
The larger the cluster, the denser the forest, and hence the more fragmented the surface plasmon.  However,
the width of the Landau fragmented surface plasmon would reach the maximum within $\Delta n=3$ transitions.  
As the cluster size further increases, the surface plasmon energy moves deeper into the forest and becomes 
degenerate with higher energy transitions (e.g. $\Delta n=5$)\cite{Yan93}, resulting in progressively 
weaker coupling and smaller width. This oscillatory behavior of the spectral width could also be understood 
in terms of the fact~\cite{MWJ02, WMWJ05} that the electron-hole density-density correlation oscillates 
as a function of cluster size due to the shell effect.

The above phenomenon can also be seen in Fig. 7, although the matrix RPA-LDA/jellium background
calculation gives the maximum width at Na$_{40}$, whereas our calculations predict it at about Na$_{65}$
(Fig. 6).  Since the present work is based on a realistic ionic background instead of the jellium model,
these transitions can no longer be characterized by $\Delta n$. However, it can clearly be seen that the
principal peak energy ($\omega_{pp}$) lies within the gap between two forests
of the electron-hole transitions for Na$_{15}$.
With the increasing cluster size, $\omega_{pp}$ comes close to and eventually falls within the right
transition forest whose edge gradually approaches the left forest at the same time.  In Na$_{27}$,
the $\omega_{pp}$ nearly touches the edge of the right transition forest,
thus resulting in the bumpy absorption profile on
the high-energy side of the principal peak (Fig. 2).  The sudden rise of the width of Na$_{51}$ (Fig. 6)
clearly indicates that the $\omega_{pp}$ has moved into the right transition forest.  For Na$_{65}$, the
$\omega_{pp}$ falls right on the position of the maximum electron-hole transitions within the right transition forest.
This results in the strongest coupling between the surface plasmon and nearly degenerate electron-hole
transitions that leads to the largest width.  Moving towards Na$_{113}$, the $\omega_{pp}$ moves further
rightward having the surface plasmon coupled with the transitions of smaller oscillator strengths, resulting
in a weaker Landau fragmentation as reflected by the narrowed width.  Afterwards, the width can be increased
or decreased, depending on the height of the unperturbed transitions the $\omega_{pp}$ encounters, but the
variation is not as large as before due to the diminishing peak height of the follow-up transitions as can
be seen from the plot for Na$_{113}$ in Fig. 7.  More measurements on the closed-shell neutral Na clusters
would still be needed to verify the present results; the discrepancy between our results and that of
Na$_{92}$ may be attributed to the missing absorptions in the experiments mentioned above (Fig. 4).
Nonetheless, some measurements on singly ionized Na clusters indeed indicated that the maximum takes place
at the clusters with 58 Na atoms or larger~\cite{RESH95, SH99}.  Thus, it is reasonable to assume that the
ionic effect would push up the cluster size in which the width maximum occurs, as Fig. 6 shows.

\section{Nature of plasmonic and electron-hole excitations}
We have just examined several electronic properties related to the optical absorption spectrum,
namely, the principal peak position, width, and static polarizability.  However, the absorption spectra
themselves do not provide further information on the electronic density behavior of plasmons or other
electronic excitations, which, however, can be made clear by referencing to the spatial distribution of
$\widetilde{\chi}(\mathbf{r},\omega)$ and the VRP spectra.  For example,
electron-hole excitations and the volume plasmon~\cite{XYK09} should both be reflected by a larger VRP value
since the induced charge density oscillations occur inside the cluster.

To analyze the nature of the prominent plasmonic resonances in the calculated absorption spectra in Figs. 3
and 5, we have evaluated the density response function to uniform perturbation,
$\widetilde{\chi}(\mathbf{r},\omega)$, of these plasmon modes.  Moreover, for studying the volume plasmon and/or
individual electron-hole excitations, several local peaks in a VRP spectrum are also analyzed.  Fig. 8
shows the $\widetilde{\chi}(\mathbf{r},\omega)$ contours on the cluster middle planes of the principal
plasmonic absorption peaks (a), and of the local peaks in the VRP spectra (b). A local peak in a VRP
spectrum indicates that
the volume mode bears a higher contribution to the oscillation charge at that frequency than at neighbouring
ones, which are featured by volume plasmon or individual electron-hole excitations.  Studying these peaks
therefore help understand the nature of these electronic excitations.

\begin{figure*}
\raggedright
(a)\begin{center}
\includegraphics[angle=0,width=16.0cm]{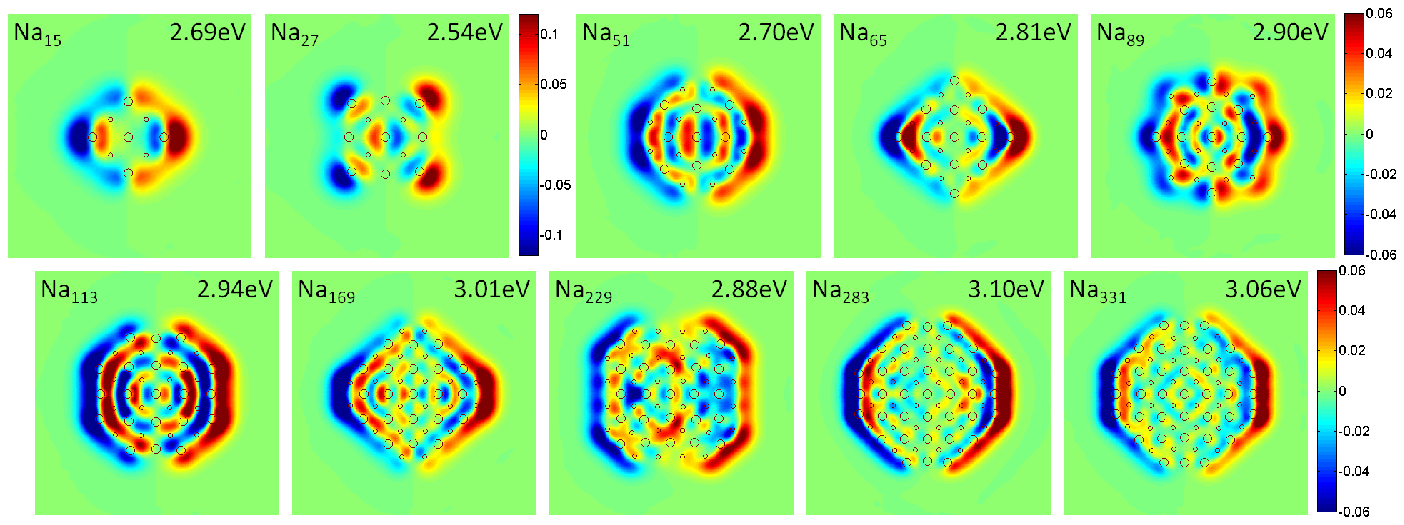}
\end{center}
(b)\begin{center}
\includegraphics[angle=0,width=16.0cm]{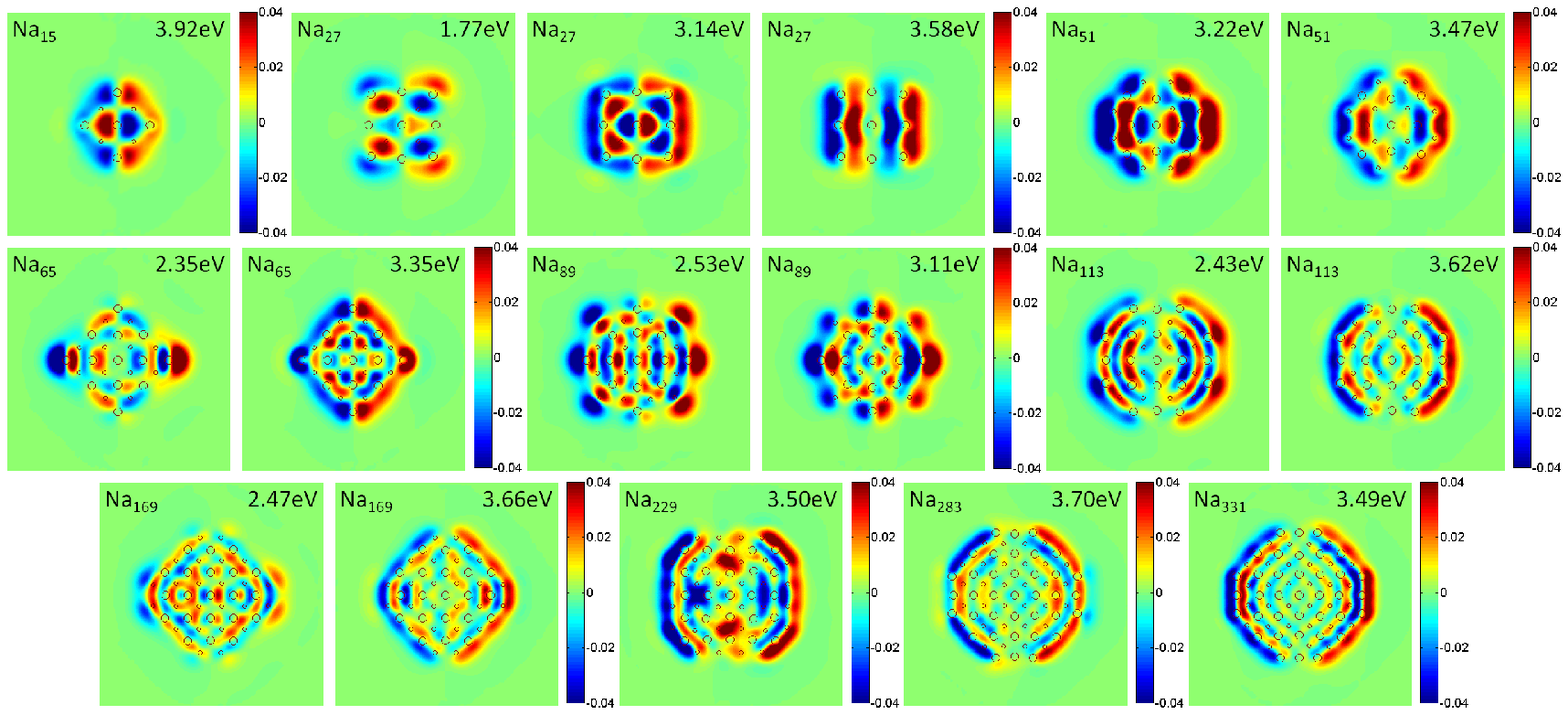}
\end{center}
(c)\begin{center}
\includegraphics[angle=0,width=16.0cm]{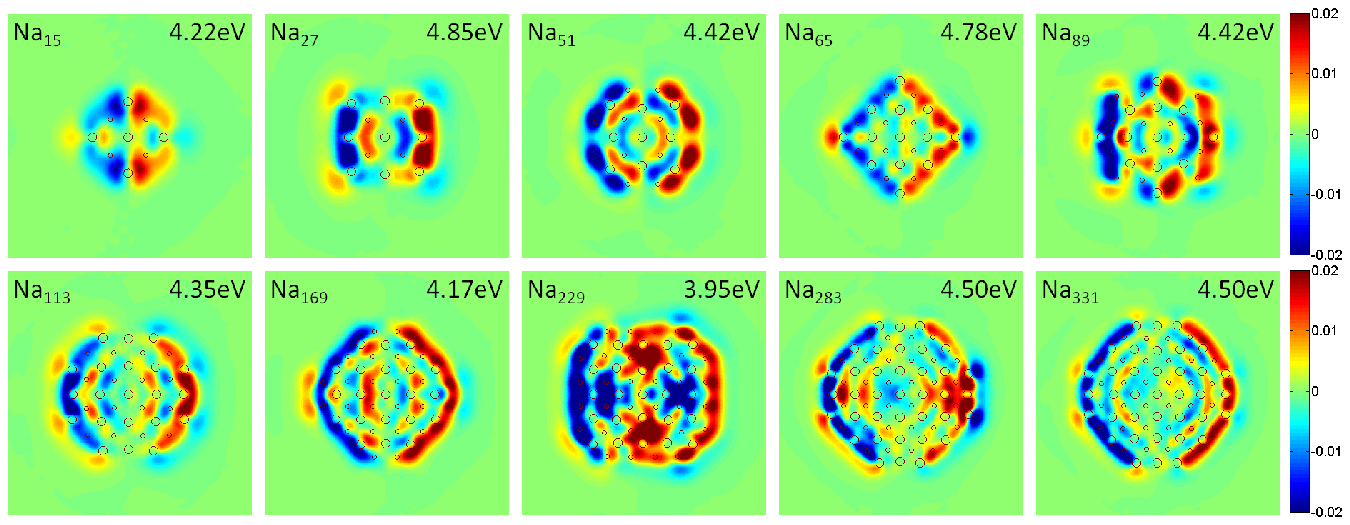}
\end{center}
\caption{(Color) Distribution of calculated linear density response function,
$\widetilde{\chi}(\mathbf{r},\omega)$, on the $xy$-plane.  The plots in panel (a) correspond to the
strongest absorption peaks, in panel (b) to the major peaks in the VRP spectra, and in panel (c) to the
local peaks in the VRP spectra around 4.5 eV on the absorption long tail.  The applied electric impulse is
along the $x$-direction.  Large and small circles, respectively, indicate atoms sitting on the $xy$-plane
and the first atoms next to the $xy$-plane.  For clarity, different contour plotting scales
have been used.  Also note that in order to match plotting scale in (a), (b), and (c), some areas of a graph
may have values exceeding its scale maximum/minimum.}
\end{figure*}

\subsection{The principal plasmon resonance}
As shown in Fig. 8a, at the principal resonance frequency, all the clusters considered here have a strong
dipolar charge cloud located outside the cluster surface, which clearly corresponds to the uniform
translational motion of electrons against the positive background in the classical Mie picture.  A similar
circumstance in nuclear physics, namely, the Goldhaber-Teller (GT) giant dipole resonance, is shown in
Fig. 9.  For most clusters the dipolar charge cloud is distributed roundly about the surface as the
Mie picture dictates, except for Na$_{27}$, which has a strong-corner signal due to the corner atoms that
make the cluster surface deviate markedly from a spherical one.  However, as the cluster size increases, this deviation
from the spherical shape due to the uneven distribution of surface atoms should become less influential,
as can be seen from the calculated charge distribution of the larger clusters in Fig. 8a.

\begin{figure}
\raggedright
\center{
\includegraphics[angle=0,width=5cm]{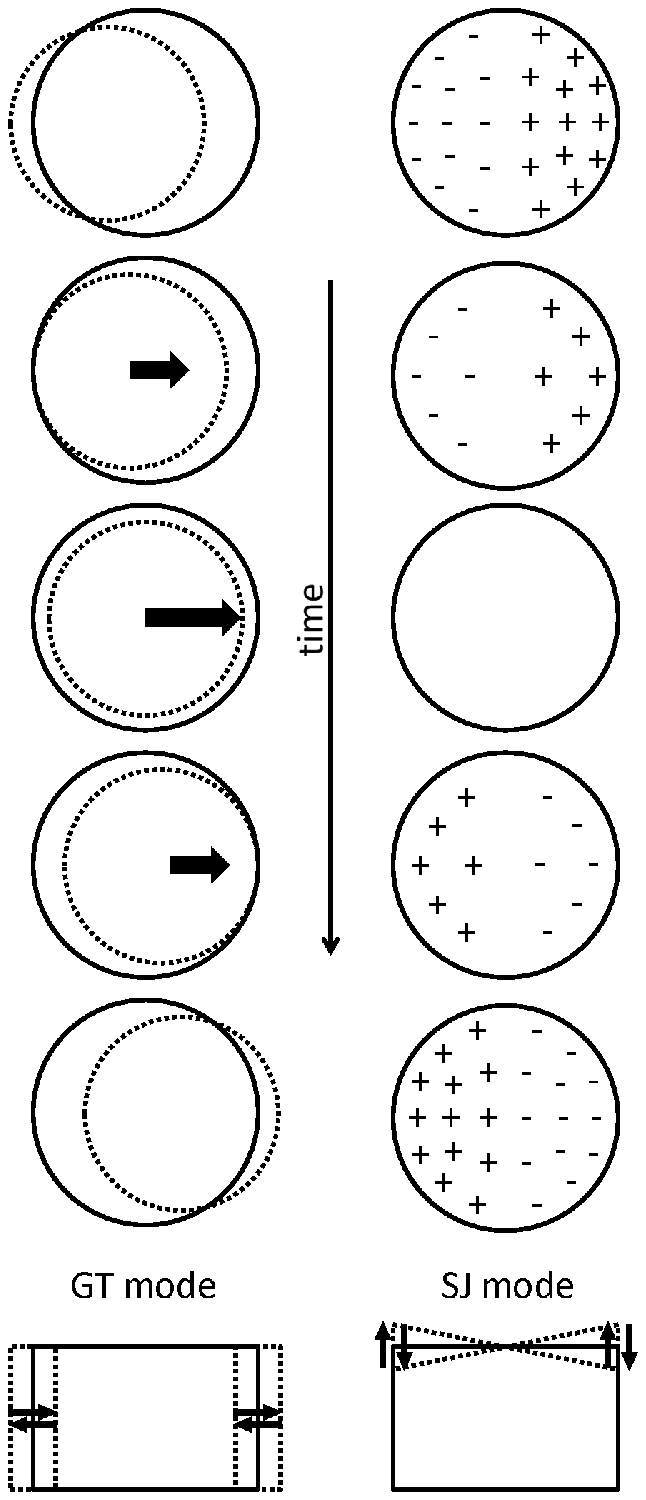}
\caption{Goldhaber-Teller (GT) and Steinwedel-Jensen-Migdal (SJ) giant resonance modes
in an atomic nucleus (adapted from
Ref.~\onlinecite{MS77}).  In the GT mode, the protons uniformly vibrate against the neutrons, whereas
in the SJ mode, the protons (and neutrons) alternately pile up at the opposite sides of a fixed boundary.
An illustration for the similar circumstances of collective oscillations of electrons in a metallic
sphere is plotted at the bottom (adapted from Ref.~\onlinecite{BWM08}).}}
\end{figure}

Careful examination of the density response contour plots in Fig. 8a reveals that the Na clusters
considered here can be divided into two groups, namely, volume mode group (called group-V) and surface
mode group (called group-S). We can see that Na$_{27}$, Na$_{51}$, Na$_{65}$, Na$_{89}$, and
Na$_{113}$ have a comparably thicker dipolar charge cloud inside their clusters that resemble several
concentric circles with alternating phases toward the cluster center, hence belonging to the group-V
clusters.  In contrast, the dipolar charge cloud inside the clusters of Na$_{15}$, Na$_{169}$, Na$_{229}$,
Na$_{283}$, and Na$_{331}$ is much weaker, and thus, these clusters are among the group-S clusters.

In consistent with the above observation, we can see in Fig. 2 that the group-S clusters have the principal
absorption peak located in a valley of the VRP spectrum, indicating a weak volume mode contribution to
the plasmonic resonance. For the group-V clusters, on the other hand, a VRP valley becomes barely
visible at, e.g., Na$_{113}$.  For a group-S cluster larger than Na$_{113}$, generally,
the principal absorption peak becomes dominated by the surface dipole mode (see Fig. 8a)
and a valley in the VRP spectrum appears near the energy of the principal absorption peak.  The only
exception is Na$_{15}$, which, despite of its small size, belongs to the group-S.  
To understand this, we resort again to Fig. 8 for a qualitative picture.  
As discussed in the preceding section, the principal
peak of Na$_{15}$ is located within the gap between the left and right electron-hole transition forests, and
hence is not Landau fragmented.  This is clearly reflected in its simple absorption profile, a small VRP
value at the principal peak energy (Fig. 2), and the weak volume mode inside the cluster (Fig. 8a).  For
Na$_{27}$, the first group-V cluster, on the other hand, the principal peak energy almost touches the right
transition forest, thus indicating the almost onset of Landau fragmentation.  The strong coupling between
the surface plasmon and the nearly degenerate electron-hole transitions therefore gives rise to a
strong mixture of the surface and volume modes for the group-V clusters (Fig. 8a).
On the other hand, for the group-S clusters Na$_{169}$, Na$_{229}$, Na$_{283}$, and
Na$_{331}$, the principal peak energy is well within the transition forest and the principal peak has the
coupling from the weaker transitions.  Since the number of transitions that are nearly degenerate with the
surface plasmon is huge and the average coupling is weaker, these numerous weak contributions would
distribute over the whole cluster with a small magnitude, as shown in Fig. 8a for Na$_{169}$, Na$_{229}$,
Na$_{283}$, and Na$_{331}$.  The progressively deeper and smoother valley in the VRP spectrum in the
vicinity of the principal absorption peak as a function of the cluster size, is regarded as the signature
for the increasing number of weaker couplings of the nearly degenerate electron-hole transitions to the
surface plasmon.

\subsection{Volume plasmon and electron-hole excitations}
Fig. 2 shows that next to the principal peak, there are several less prominent peaks as well,
especially for the group-V clusters.  As discussed before, these peaks result predominantly from the Landau
fragmentation from the coupling of the surface plasmon to the nearly degenerate electron-hole transitions.
Because these excitations retain most of their electron-hole transition nature, a large density
response inside the cluster is anticipated, which is reflected by a large VRP value.  Therefore, 
the VRP spectrum may be used to locate these electronic excitations that would otherwise be unrecognizable 
in the absorption spectrum due to their small transition dipole moment.

Fig. 8b displays the $\widetilde{\chi}(\mathbf{r},\omega)$ of several VRP peaks in the vicinity of the
principal absorption peak for each cluster.  As shown in Fig. 2, there are two VRP peaks for Na$_{15}$ near
4 eV, which should correspond to two electron-hole transitions that are not yet strongly coupled with the
surface plasmon.  For Na$_{27}$ and larger clusters, on the other hand, the strong coupling should occur
involving the descendent of the first VRP peak of Na$_{15}$ at 3.92 eV.  For example, it can be seen that
the $\widetilde{\chi}(\mathbf{r},\omega)$ plot of Na$_{27}$ at 3.14 eV strongly resembles (but in the
opposite phase with) that of Na$_{15}$ at 3.92 eV in the same spatial area of the two clusters.  After
Landau fragmentation, the resultant electron-hole excitations exhibit various complicated spatial
distributions (Fig. 8b), which are fostered by the discrete ionic background and the uneven cluster surface.
The intense contour magnitudes both inside and on the surface of the group-V clusters are the consequence of
the strong coupling between the surface plasmon and the nearly degenerate electron-hole transitions.  For
the group-S clusters of greater size, on the other hand, the number of nearly degenerate transitions
increases whereas the average strength of each coupling reduces, resulting in the reduced maximum of
$\widetilde{\chi}(\mathbf{r},\omega)$ inside the clusters.

Interestingly, the second VRP peak at 4.22 eV of Na$_{15}$ evolves into the VRP peaks near 4.5 eV
of larger clusters, as evidenced by the similar magnitude and distribution of
$\widetilde{\chi}(\mathbf{r},\omega)$ shown in Fig. 8c.  This VRP peak corresponds to the absorption
long tail extending to $\sim$5.5 eV that can be seen in Fig. 2.  This long tail in the UV range was
confirmed by very recent experiments~\cite{XYK09} (Fig. 4) and was assigned to the volume plasmon that can
be optically excited for finite particles as predicted by the previous theoretical
works~\cite{Kre90, Bra93, Eka85, Bec91, Yan93, MG95, RA97, RBR09}.  The surface and volume plasmons can be
paralleled, respectively, to the Goldhaber-Teller (GT) and Steinwedel-Jensen-Migdal (SJ) modes of giant
dipole resonances in an atomic nucleus, as depicted in Fig. 9.  Evidently, the diagrams shown in Fig. 8c for
the absorption long tail of each cluster correspond to the SJ mode for which electrons are piled up
alternatingly at the opposite sides of the cluster, despite that the ionic background and rough cluster
surface may be responsible for the VRP spectrum fluctuations and the microscopic density response features
as exhibited by these contours.

Overall, as can be expected, the energy, shape, and the density response nature of the principal absorption
peak approach the predictions of the classical Mie theory as the cluster size increases.  However,
while the matrix RPA-LDA/jellium background calculations~\cite{Yan93} predicted that a smooth principal
peak starts to take shape at around Na$_{58}$, the present work shows that this happens only after
Na$_{113}$, in company with the formation of a valley in the VRP spectrum (Fig. 2) and also a density
response similar to that of the Mie picture (Fig. 8a), namely, a uniform translational motion of electrons
against the positive background (Fig. 9).  The Na$_{113}$ cluster has an effective diameter of $\sim$2.0 nm.

While the optical absorption of the metallic clusters is anticipated to become dominated by the surface
plasmon as the cluster approaches the bulk limit~\cite{Bra93}, our results nevertheless indicate that the
prominent asymmetric line shape of the principal peak in the absorption spectrum remains even for clusters
as large as Na$_{331}$, as evidenced by its additional absorptions in the UV range (Fig. 2).  It is clear
from the VRP spectrum that this absorption structure consists of the volume plasmon ($\sim$4.5 eV) and also
the excitations formed by the coupling between the surface plasmon and the electron-hole transitions.

\section{Conclusions}
In this paper, we have studied the electronic excitation properties of a series of sphere-like bcc Na
clusters up to a size that is not yet explored in literature, by real time TDLDA calculations. Several
properties related to the optical absorption spectrum, namely, the principal peak position and width as well
as static polarizability, have been presented.  Moreover, the calculated VRP spectrum (i.e. volume density
response proportion at different frequencies) has been used to locate the volume plasmon and electron-hole
excitations with weak transition dipole moments. The spatial distribution of density response function for
these special frequencies has also been reported. The good agreement in the detailed spectral features
between Na$_{27}$ (Na$_{89}$) (our calculations) and Na$_{20}$ (Na$_{92}$) (previous
experiments) as well as in the static polarizability between the present work and previous experiments
suggests that our TDLDA calculations that take into account the realistic ionic background would provide an
adequate description of the plasmonic excitations in quantum-sized Na clusters considered here.

Firstly, we have demonstrated that the effect of the ionic background (i.e., ionic species and lattice)
is responsible for the remaining discrepancy in the principal absorption peak energy between the experiments
and previous TDLDA/jellium background calculations.  In other words, taking all the electron spill-out
effect, the coupling between the individual electron-hole transitions and surface plasmon, and the ionic
background influence into account would satisfactorily explain the observed deviation of the principal peak
energy of finite size clusters from the classical Mie theory.

Secondly, we find that the ionic background effect would push the critical cluster size where the maximum
width of the principal peak occurs from Na$_{40}$ predicted by the matrix RPA-LDA/jellium background
calculation~\cite{Yan93} to Na$_{65}$.  This is determined by the strength of coupling between the surface
plasmon and nearly degenerate electron-hole transitions that is the mechanism of Landau fragmentation. The
strong coupling also gives rise to a multiple absorption peak structure near the principal peak in the
group-V clusters, namely, Na$_{27}$, Na$_{51}$, Na$_{65}$, Na$_{89}$, and Na$_{113}$. As a result, the
density response function for the principal and VRP peaks of these clusters is dominated by an intense
volume mode inside the clusters.  On the other hand, the group-S clusters excluding Na$_{15}$ (i.e.,
Na$_{169}$, Na$_{229}$, Na$_{283}$, and Na$_{331}$) exhibit a smoother and narrower principal absorption
peak and a VRP valley near the principal peak because their surface plasmon energy is located deeply among
that of the unperturbed electron-hole transitions with weaker oscillator strengths. Their density responses
already bear resemblance to that of the classical Mie theory.

Finally, we have attributed the absorption long tail in the UV range to the volume plasmon that exists only
in finite particles with the charge density response paralleled to the giant SJ resonance mode in an atomic
nucleus.  This volume plasmon manifests itself in the absorption spectrum even for clusters as large as
Na$_{331}$, and hence cannot be ignored.  Indeed, finite size metal clusters are found to exhibit quite
complicated electronic excitations due to their quantum size and discrete ionic background.  
Although the volume plasmon characteristics may vary rather smoothly with the
increasing cluster size, the surface plasmon is strongly modified by the Landau fragmentation and the
classical features become dominant only when the clusters are larger than Na$_{113}$ which has an effective
diameter of $\sim$2.0 nm.

\section*{Acknowledgements}
We gratefully acknowledge supports from the National Science Council and the National Center for Theoretical
Sciences of Taiwan as well as the Center for Quantum Science and Engineering, National Taiwan University
(CQSE-10R1004021). We also thank the National Center for High-performance Computing of Taiwan for providing
CPU time.

\bibliographystyle{unsrt}
\bibliography{testbib}

\begin{thebibliography}{99}

\bibitem{Mai07} S. A. Maier, Plasmonics: Fundamentals and Applications, Springer, 2007.
\bibitem{ZPN09} J. Zuloaga, E. Prodan, and P. Nordlander, Nano Lett. \textbf{9}, 887 (2009).
\bibitem{HLC11} N. J. Halas, S. Lal, W. S. Chang, S. Link, and P. Nordlander, Chem. Rev. \textbf{111}, 3913 (2011).
\bibitem{MdN11} A. Manjavacas, F. J. G. de Abajo, and P. Nordlander, Nano Lett. \textbf{11}, 2318 (2011).
\bibitem{SKD12} J. A. Scholl, A. L. Koh, and J. A. Dionne, Nature \textbf{483}, 421 (2012).
\bibitem{Mie8} G. Mie, Leipzig, Ann. Phys. \textbf{330}, 377 (1908).
\bibitem{Fer58} R. A. Ferrell, Phys. Rev. \textbf{111}, 1214 (1958).
\bibitem{Bra93} M. Brack, Rev. Mod. Phys. \textbf{65}, 677 (1993).
\bibitem{BWM08} M. Brack, P. Winkler, and M. V. N. Murthy, Int. J. Mod. Phys. E \textbf{17}, 138 (2008).
\bibitem{Kre90} V. Kresin, Phys. Rev. B \textbf{42}, 3247 (1990).
\bibitem{XYK09} C. Xia, C. Yin, and V. V. Kresin, Phys. Rev. Lett. \textbf{102}, 156802 (2009).
\bibitem{YG08} J. Yan and S. Gao, Phys. Rev. B \textbf{78}, 235413 (2008).
\bibitem{NKR02} V. O. Nesterenko, W. Kleinig, and P.-G. Reinhard, Eur. Phys. J. D \textbf{19}, 57 (2002).
\bibitem{Yan93} C. Yannouleas, E. Vigezzi, and R. A. Broglia, Phys. Rev. B \textbf{47}, 9849 (1993).
\bibitem{Yan98} C. Yannouleas, Phys. Rev. B \textbf{58}, 6748 (1998).
\bibitem{MUN06} M. A. L. Marques, C. A. Ullrich, F. Nogueira, A. Rubio, K. Burke, E. K. U. Gross (Eds.), Time-Dependent Density Functional Theory, Springer-Verlag, 2006.
\bibitem{RG84} E. Runge, E. K. U. Gross, Phys. Rev. Lett. \textbf{52}, 997 (1984).
\bibitem{Sla74} J. C. Slater, in: The Self-Consistent Field for Molecular and Solids, Quantum Theory of Molecular and Solids, vol. 4, McGraw-Hill, New York, 1974.
\bibitem{VWN80} S. H. Vosko, L. Wilk, M. Nusair, Can. J. Phys. \textbf{58}, 1200 (1980).
\bibitem{JC72} P. B. Johnson and R. W. Christy, Phys. Rev. B \textbf{6}, 4370 (1972).
\bibitem{PNH03} E. Prodan, P. Nordlander, and N. J. Halas, Chem. Phys. Lett. \textbf{368}, 94 (2003).
\bibitem{YB96} K. Yabana and G. F. Bertsch, Phys. Rev. B \textbf{54}, 4484 (1996).
\bibitem{ZS80} A. Zangwill and P. Soven, Phys. Rev. A \textbf{21}, 1561 (1980).
\bibitem{Eka85} W. Ekardt, Phys. Rev. B \textbf{31}, 6360 (1985).
\bibitem{C95} M. E. Casida, in: D. P. Chong (Ed.), Recent Advances in Density Functional Methods, vol. 1, World Scientific, Singapore, 1995.
\bibitem{dH93} W. A. de Heer, Rev. Mod. Phys. \textbf{65}, 611 (1993).
\bibitem{Kni84} W. D. Knight, K. Clemenger, W. A. de Heer, W. A. Saunders, M. Y. Chou, and M. K. Cohen, Phys. Rev. Lett. \textbf{52}, 2141 (1984).
\bibitem{Bra89} M. Brack, Phys. Rev. B, \textbf{39}, 3533 (1989).
\bibitem{LRM91} G. Lauritsch, P.-G. Reinhard, J. Meyer, and M. Brack, Phys. Lett. A \textbf{160}, 179 (1991).
\bibitem{MR95} B. Montag and P.-G. Reinhard, Phys. Rev. B \textbf{51}, 14686 (1995).
\bibitem{Bec91} D. E. Beck, Phys. Rev. B \textbf{43}, 7301 (1991).
\bibitem{MG95} M. Madjet, C. Guet, and W. R. Johnson, Phys. Rev. A \textbf{51}, 1327 (1995).
\bibitem{CRS98} F. Calvayrac, P.-G. Reinhard, and E. Suraud, J. Phys. B: At. Mol. Opt. Phys. \textbf{31}, 1367 (1998).
\bibitem{WIJ06} G. Weick, G.-L. Ingold, R. A. Jalabert, and D. Weinmann, Phys. Rev. B \textbf{74}, 165421 (2006).
\bibitem{HB04} P.-A. Hervieux and J.-Y. Bigot, Phys. Rev. Lett. \textbf{92}, 197402 (2004).
\bibitem{SEK99} M. Schmidt, C. Ellert, W. Kronm\"{u}ller, and H. Haberland, Phys. Rev. B \textbf{59}, 10970 (1999).
\bibitem{SSG02} I. A. Solov'yov, A. V. Solov'yov, and W. Greiner, Phys. Rev. A \textbf{65}, 053203 (2002).
\bibitem{KBR00} S. K\"{u}mmel, M. Brack, and P.-G. Reinhard, Phys. Rev. B \textbf{62}, 7602 (2000).
\bibitem{RS04} P.-G. Reinhard and E. Suraud, Introduction to Cluster Dynamics, Wiley-VCH, 2004.
\bibitem{YL94} C. Yannouleas and U. Landman, Phys. Rev. B \textbf{51}, 1902 (1995).
\bibitem{AM76} N. W. Ashcroft and N. D. Mermin, Solid State Physics, Brooks Cole, 1976.
\bibitem{PBE96} J. P. Perdew, K. Burke, and M. Ernzerhof. Phys. Rev. Lett. \textbf{77}, 3865 (1996).
\bibitem{KH9394} G. Kresse and J. Hafner, Phys. Rev. B \textbf{47}, 558 (1993); ibid. \textbf{49}, 14251 (1994).
\bibitem{KF96} G. Kresse and J. Furthm\"{u}ller, Comput. Mat. Sci. \textbf{6}, 15 (1996).
\bibitem{KFu96} G. Kresse and J. Furthm\"{u}ller, Phys. Rev. B \textbf{54}, 11169 (1996).
\bibitem{MCB03} M. A. L. Marques, A. Castro, G. F. Bertsch, and A. Rubio, Comput. Phys. Commun. \textbf{151}, 60 (2003).
\bibitem{TM91} N. Troullier and J. L. Martins, Phys. Rev. B \textbf{43}, 1993 (1991).
\bibitem{KB82} L. Kleinman and D. M. Bylander, Phys. Rev. Lett. \textbf{48}, 1425 (1982).
\bibitem{ARS08} K. Andrea, P.-G. Reinhard, and E. Suraud, arXiv:physics/0701028.
\bibitem{BLM79} O. Bohigas, A. M. Lane, and J. Martorell, Phys. Rep. \textbf{51}, 267 (1979).
\bibitem{SK91} K. Selby, V. Kresin, J. Masui, M. Vollmer, W. A. de Heer, A. Scheidemann, and W. D. Knight, Phys. Rev. B \textbf{43}, 4565 (1991).
\bibitem{RESH95} T. Reiners, C. Ellert, M. Schmidt, and H. Haberland, Phys. Rev. Lett. \textbf{74}, 1558 (1995).
\bibitem{GG02} L. G. Gerchikov, C. Guet, and A. N. Ipatov, Phys. Rev. A \textbf{66}, 053202 (2002).
\bibitem{PWK91} S. Pollack, C. R. C. Wang, and M. M. Kappes, J. Chem. Phys. \textbf{94}, 2496 (1991).
\bibitem{ABMR07} X. Andrade, S. Botti, M. A. L. Marques, and A. Rubio, J. Chem. Phys. \textbf{126}, 184106 (2007).
\bibitem{TK01} G. Tikhonov, V. Kasperovich, K. Wong, and V. V. Kresin, Phys. Rev. A \textbf{64}, 063202 (2001).
\bibitem{RA99} D. Rayane, A. R. Allouche, E. Benichou, R. Antoine, M. Aubert-Fercon, Ph. Dugourd, M. Broyer, C. Ristori, F. Chandezon, B. A. Huber, and C. Guet, Eur. Phys. J. D \textbf{9}, 243 (1999).
\bibitem{KAM00} S. K\"{u}mmel, J. Akola, and M. Manninen, Phys. Rev. Lett. \textbf{84}, 3827 (2000).
\bibitem{BGZ00} S. A. Blundell, C. Guet, and R. R. Zope, Phys. Rev. Lett. \textbf{84}, 4826 (2000).
\bibitem{MWJ02} R. A. Molina, D. Weinmann, and R. A. Jalabert, Phys. Rev. B \textbf{65}, 155427 (2002).
\bibitem{WMWJ05} G. Weick, R. A. Molina, D. Weinmann, and R. A. Jalabert, Phys. Rev. B \textbf{72}, 115410 (2005).
\bibitem{SH99} M. Schmidt and H. Haberland, Eur. Phys. J. D \textbf{6}, 109- (1999).
\bibitem{MS77} W. D. Myers, W. J. Swiatecki, T. Kodama, L. J. El-Jaick, and E. R. Hilf, Phys. Rev. C \textbf{15}, 2032 (1977).
\bibitem{RA97} A. Rubio, J. A. Alonso, X. Blase, and S. G. Louie, Int. J. Mod. Phys. B \textbf{11}, 2727 (1997).
\bibitem{RBR09} A. A. Raduta, R. Budaca, and Al. H. Raduta, Phys. Rev. A \textbf{79}, 023202 (2009).

\end{thebibliography}

\end{document}